\documentclass{IEEEcsmag}

\usepackage[colorlinks,urlcolor=blue,linkcolor=blue,citecolor=blue]{hyperref}
\expandafter\def\expandafter\UrlBreaks\expandafter{\UrlBreaks\do\/\do\*\do\-\do\~\do\'\do\"\do\-}
\usepackage{upmath,color}
\usepackage{algorithmicx}
\usepackage{booktabs}
\usepackage{longtable}
\usepackage{stmaryrd}
\usepackage{xspace}
\usepackage[numbers,sort&compress]{natbib}

\jvol{XX}
\jnum{XX}
\paper{8}
\jmonth{March}
\jname{Computing in Science \& Engineering}
\jtitle{A pragmatic workflow for research software engineering in computational science}
\pubyear{2023}

\newcommand{\dg}[1]{\color{black}#1\color{black}\xspace}

\newcommand{\eg}{e.g.\@\xspace}

\newcommand{\minwf}{minimal workflow}

\usepackage[final,commentmarkup=uwave]{changes} % defaultcolor=blue
\definechangesauthor[name=dennis, color=green]{dg}
\definechangesauthor[name=tomislav, color=cyan]{tm}
\usepackage{cleveref}

\begin{document}

%\sptitle{A Research Software Engineering Workflow for Computational Science and Engineering}
% \title{A Research Software Engineering Workflow for Computational Science and Engineering}

\title{A pragmatic workflow for research software engineering in computational science}
\sptitle{A pragmatic workflow for research software engineering in computational science}

\author{Tomislav Maric}
\affil{Mathematical modeling and analysis, Mathematics department, TU Darmstadt, Darmstadt, 64287, Germany\footnote{maric@mma.tu-darmstadt.de}}

\author{Dennis Gl\"{a}ser}
\affil{Institute for Modelling Hydraulic and Environmental Systems, Department of Hydromechanics and Modelling of Hydrosystems, University of Stuttgart, Stuttgart, 70569, Germany\footnote{dennis.glaeser@iws.uni-stuttgart.de}}

\author{Jan-Patrick Lehr}
\affil{Institute for Scientific Computing, Department of Computer Science, TU Darmstadt, Darmstadt, 64289, Germany\footnote{jan-patrick.lehr@tu-darmstadt.de}}

\author{Ioannis Papagiannidis}
\affil{Mathematical modeling and analysis, Mathematics department, TU Darmstadt, Darmstadt, 64287, Germany\footnote{TODO}}

\author{Benjamin Lambie}
\affil{Research Field Energy \& Environment, TU Darmstadt, Darmstadt, 64287, Germany\footnote{lambie@tfi.tu-darmstadt.de}}

\author{Christian Bischof}
\affil{Institute for Scientific Computing, Department of Computer Science, TU Darmstadt, Darmstadt, 64289, Germany\footnote{christian.bischof@tu-darmstadt.de}}

\author{Dieter Bothe}
\affil{Mathematical modeling and analysis, Mathematics department, TU Darmstadt, Darmstadt, 64287, Germany\footnote{bothe@mma.tu-darmstadt.de}}

%\markboth{THEME/FEATURE/DEPARTMENT}{THEME/FEATURE/DEPARTMENT}

\begin{abstract}\looseness-1University research groups in Computational Science and Engineering (CSE) lack funding and personnel for Research Software Engineering (RSE), which leads to a focus on journal publications over sustainable software development and reproducible results. This negatively impacts CSE research output, with data and software related to publications being difficult to find, reproduce, or reuse. We propose an RSE workflow for CSE that addresses these challenges and improves research output quality by applying established software engineering practices such as version control and testing, as well as cross-linking software with data and publications. The workflow is minimal in terms of overhead and produces tested software cross-linked with publications and reproducible results. 
We define research software quality as the ability to quickly find and understand related publications, data, and software, and to extend existing code to realize new research ideas.
%\comment[id=dg]{Does the workflow adress modularity, really -> and thus, the ability to extend existing code? My impression is that modularity/extensibility are a quality on the source code level, which is below what our workflow addresses? Let's discuss this, depending on the outcome we may have to modify this last phrase of the abstract.}
%\comment[id=tm]{Great point! Nope, it doesn't, in any way.}
%\comment[id=dg]{We should harmonize citing throughout the document. For instance, inline links or footnotes should probably be removed and substituted by citing an entry in the bibliography for those websites}
%\comment[id=tm]{We are limited to 12 citations, we must cite everything else as a footnote, see TODOs, you mean remove duplicates?}
\end{abstract}

\maketitle

\chapteri{R}esearch software \dg{engineering} - crucial in Computational Science and Engineering (CSE) - continues to find little application in academia, potentially due to beliefs that the costs outweigh the benefits~\citep{2020:heroux:psip}. However, efficient and long-term CSE research is impossible without sustainable research software development, and it is important that researchers are able to \textbf{F}ind, \textbf{A}ccess, \textbf{I}nteract with and \textbf{R}euse research software, while being able to \textbf{R}eproduce results from publications (\textbf{FAIR} \citep{FAIR}).

Reproducibility is key to transparent and trustworthy research. To achieve this, it is crucial to have access to the research software and its dependencies, as well as the configuration and input data used in a publication. Ideally, an automated description of the entire computational pipeline is available in order to simplify the reproduction of published results. Without access to the resources, peers have to re-implement the presented methods, which often requires knowledge about details that are not reported in journal publications. Besides this, re-implementation can significantly slow down the research process, effectively leading to financial losses in public research funding.

In recent years, sustainable research software development has gained a lot of traction in the CSE community. Initiatives such as Better Scientific Software (BSS\footnote{\url{https://bssw.io}}), the National Research Data Initiative (NFDI\footnote{\url{https://www.nfdi.de}})) (with a community-driven knowledge-base\footnote{\url{https://nfdi4ing.pages.rwth-aachen.de/knowledge-base/}}), and the Sustainable Software Initiative\footnote{\url{https://software.ac.uk}}, support sustainable development by organizing workshops and providing best practices to the CSE community. Another community-driven source of information on increasing the reproducibility of research results in data science is the ``Turing Way'' handbook\footnote{\url{https://the-turing-way.netlify.app/welcome.html}}.

The issue of sustainable research software development was also discussed by several authors in the literature. \citet{Wilson2014} present best practices for increasing the quality of research software in terms of consistent naming and formatting, quality assurance measures (automated build and test pipelines), design principles (\eg ``don't-repeat-yourself'') and the use of version control systems. \citet{stanisic_effective_2015} propose a git-workflow\footnote{\url{https://git-scm.com}} that includes research data and \textit{Org-mode}\footnote{\url{https://orgmode.org}} for documentation in the Emacs text editor. Continuous Integration (CI) for research software on HPC systems has been proposed in \citep{sampedro_continuous_2018}, using Jenkins CI servers and Singularity (Apptainer) containers\footnote{\url{http://apptainer.org}}. This approach is promising, particularly the encapsulation of research software in containerized environments. However, current web-based version control services like GitHub and GitLab offer integrated CI services, rendering dedicated servers unnecessary for small research groups. 

Other authors focus on ``good enough practices'' in scientific computing~\citep{wilson_good_2017}, comprising code quality in terms of naming and comments, archival of easily understandable raw data, as well as recommendations on project organization. ``Good enough'' is a crucial attribute for small university research groups: a workflow must be simple enough to be quickly adopted by a small team of Ph.D. students and postdocs. They may further be motivated to lay more focus on code quality by making the code publicly accessible early on. Early public access, the use of a compatible license, and having transparent contribution and communication channels are important for open-source projects that aim to attract collaboration partners \cite{Jimenez2017}.

Our proposed workflow targets small to medium-sized research groups at Universities, with the goal to continuously yield \textit{FAIR} research software and related research results. The workflow applies established techniques from Software Engineering to research software in the context of scientific publications. Acknowledging very limited resources for software engineering at universities, we propose a minimal and a full workflow. \nameref{ssec:vcs} connects version control with the peer-review process for scientific publications, to ensure integration, traceability and re-use of research ideas from publications. \nameref{ssec:cross-linking} makes digital research artefacts findable and reproducibile. The full workflow consists of more advanced techniques: \nameref{ssec:tdd} focuses first and foremost on tests whose results are published in scientific publications, and adds new tests in a top-down approach. \nameref{ssec:tests} delivers secondary data that plays a crucial role in the peer-review process, in effect determining the quality of a research idea and the acceptance of a publication. Furthermore, secondary data presents the basis for comparison of different research ideas - we propose a pragmatic, straightforward, open and inter-operable format for secondary data and metadata. \nameref{ssec:containerization} is used to encapsulate research software - without tailoring containers to different large-scale HPC systems. \nameref{ssec:continuous-integration} utilises tests that are added in the process of generating scientific publications, by lowering the resolution of computational problem such that they are affordable to run within CI pipelines. 
\Cref{table:comparison} summarizes the comparison of our workflow with other workflows reported in the literature.

\begin{table*}[!hbt]
            \centering
		\begin{tabular}{@{} *6l @{}}    \toprule
				\emph{Publication} & \emph{Branching model} & \emph{Test-Driven Development} & \emph{Cross-linking} & \emph{Continuous Integration}  & Secondary data \\\midrule
				 \citep{Jimenez2017}
					 & -  & -  & -  & - & 1  \\ 
				 \citep{Fehr2016} 
					 & -  & -  & -  & - & 2  \\ 
				 \citep{Wilson2014} 
					 & 1  & 2  & -  & - & -  \\ 
				 \citep{wilson_good_2017}
					 & -  & -  & 3 & 1 & 3  \\ 
				 \citep{stanisic_effective_2015}
					 & 1  & -  & - & 1 & -  \\ 
				 \citep{anzt_towards_2019}
					 & 1  & -  & - & 5 & -  \\ 
				 \citep{riesch_bertha_2020}
					 & 1  & -  & - & 1 & 4  \\ 
				 \citep{sampedro_continuous_2018}
					 & 1  & -  & -  & 4 & - \\
					 \bottomrule
				 \hline
		\end{tabular}
            \medskip
		\caption{A comparison of the proposed workflow with other published workflows. Similarity is expressed using numbers 1 (not similar) to 5 (very similar), or '-' for an aspect that is not addressed.}
		\label{table:comparison}
\end{table*}
\section{RESEARCH SOFTWARE ENGINEERING WORKFLOW}
\label{sec:workflow}

\begin{figure}[tb]
    \centering
    \includegraphics[clip=true,trim={1.3cm 20.5cm 3cm 0.1cm},width=\linewidth]{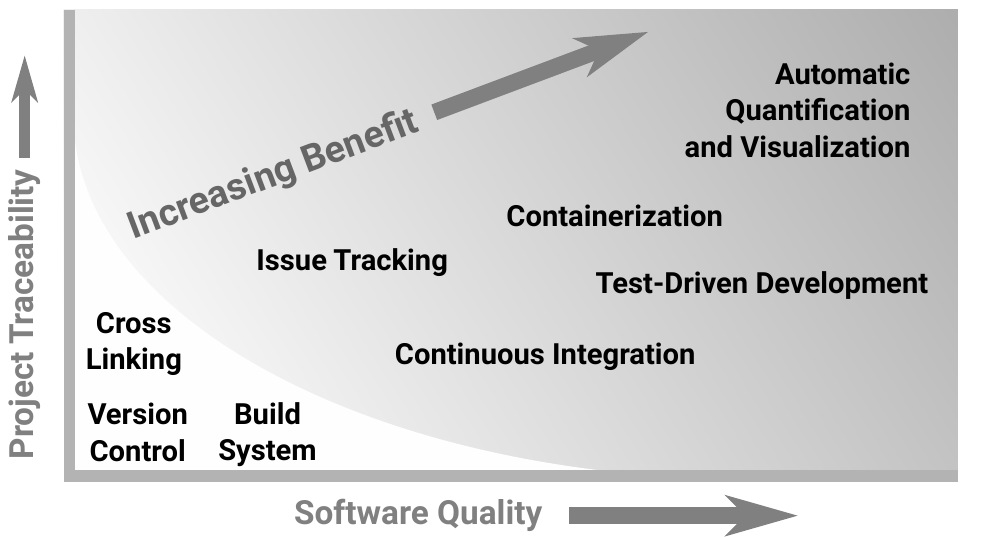}
    \caption{Impact of the proposed workflow on software quality and project traceability. Elements in the lower white area show the \minwf{} steps. %The elements along the gradient are recommended elements that comprise the \fuwf{}.
    }
    \label{fig:workflow-overview}
\end{figure}

\Cref{fig:workflow-overview} highlights the most relevant aspects of the minimal version of our workflow. The minimal workflow combines a version control system for all source files and experiment configurations, an established build system, with linking digital research artifacts (code, publications, and data) with persistent identifier (PIDs) from data repositories.
% This approach introduces a very small work overhead, while providing significant benefits.
%
%Additions to the minimal workflow involve issue tracking, continuous integration, test-driven development, containerization, and automatic result visualization.}Projects should establish the minimum aspects first, and implement further aspects in a gradual manner~\citep{2020:heroux:psip}.

% Simplified workflow part
\subsection{MINIMAL WORKFLOW}

\label{sec:minimal-workflow-short}

\subsection{Version-control branching-model for research software}
\label{ssec:vcs}

\begin{figure*}[htb]
    \centering
    \includegraphics[width=\linewidth]{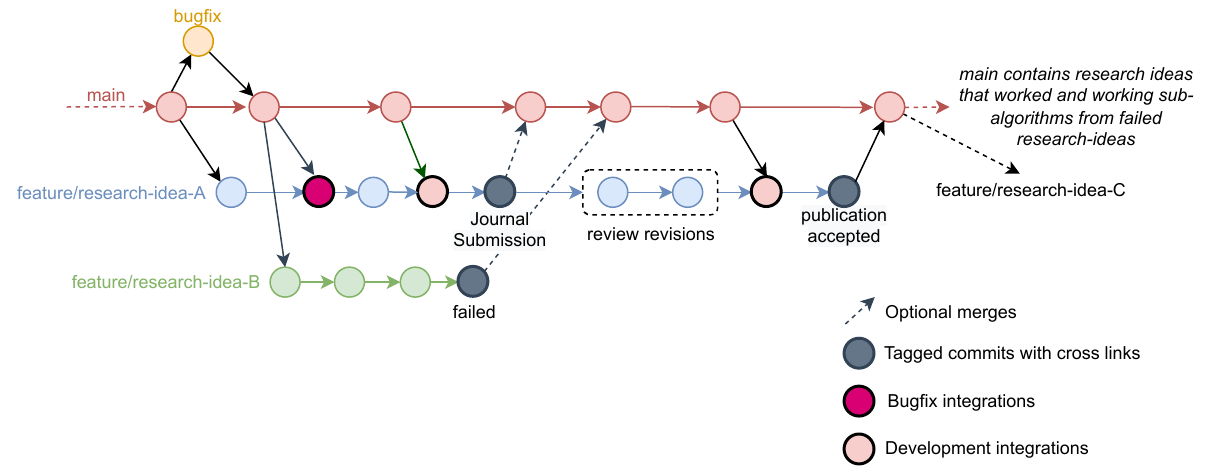}
    \caption{Schematic of the relationship between the feature-based git branching model, the peer-review process, and the cross-linking of digital research artifacts. The individual features required for a research idea are depicted as individual commits for the sake of readability, but they may be developed on dedicated branches.}
    \label{fig:branching}
\end{figure*}

Frequent personnel changes make version control essential in the university research environment. In particular, Ph.D. students need to be able to efficiently and effectively re-use previous research results.

The proposed workflow assumes that a research group develops and extends a code base within different ongoing research projects. Furthermore, note that we use concepts and terms related to "git" - a well-established version-control system.
%{The proposed workflow assumes that a research group develops a code base and extends it within different ongoing research projects. Furthermore, note that for the description of the workflow, we use concepts and terms of the well-established version-control system (VCS) \emph{git} which is well-established, and the many web interfaces such as Gitlab~\citep{GitLab}, Github~\citep{GitHub}, or Bitbucket~\citep{Bitbucket} simplify its use.}

We use the term \textit{research idea} to refer to a unit of development that is anticipated to lead to a journal publication. Such ideas may require many new features, generally developed in parallel on different branches. However, we suggest creating one branch for each research idea, that will contain all information needed for a scientific publication. With a focus on small research groups, we suggest using a feature-based branching model\footnote{\url{https://nvie.com/posts/a-successful-git-branching-model}} that is closely related to the scientific publication process, as shown schematically in \cref{fig:branching}.

A research idea branch can be closed once all required features are incorporated and a) the delivered results have publication-level quality, or b) the research idea fails. In the former case, the results that the software produces in the current state may be discussed in a preprint and submitted to a scientific journal for a peer-review process. In the latter case, writing a report about a negative scientific result is still very beneficial, especially since Ph.D. researchers generally leave the institution.

In either case, before writing up the report, testing existing functionality (features) is necessary to ensure that the new developments did not introduce errors in other parts of the code base, and that new results constitute an improvement. Ideally, the testing procedures are automated. At this point, we suggest that a snapshot of the research software, related data (e.g. input or output data) and the report are published on dedicated data repositories or preprint servers such as Zenodo\footnote{\url{https://zenodo.org}}, TUdatalib\footnote{\url{https://tudatalib.ulb.tu-darmstadt.de}}, or ArXiv\footnote{\url{https://arxiv.org}}. This yields Persistent Identifiers (PIDs) that can be used for cross-linking (cf. \nameref{ssec:cross-linking}). Reports of negative findings may also be published on a university-internal server. We suggest creating the datasets for code and data before publishing the report, such that those PIDs can be added to the report itself before publishing it.

All PIDs may then be mentioned in the README file of the source code, a change introduced to the code repository by a dedicated \textit{commit}. Afterwards, this commit can be \textit{tagged} following a naming convention established in the research group (e.g. \textit{year-journal-topic-revision}), and this tag is explicitly referenced in the metadata of the source code publication mentioned above. 

At this point, persistent versions of the code, related data and textual documentation exist and are cross-linked (cf. \nameref{ssec:cross-linking}) to a specific version in the actively developed code repository via the \textit{git tag}. The research group may decide to merge the research idea branch although the peer-review process is still ongoing, for instance, if preexisting features have been improved on the research idea branch.

Development of the research idea branch continues throughout the peer-review process. Review comments may lead to new feature branches and bug fixes, which are all merged before producing new results. The datasets mentioned above are updated by newer versions for each revision and cross-linked again. Finally, the research idea branch is merged into the main branch when a publication is accepted.

Therefore, the primary (tagged) milestones in this git branching model revolve around scientific publications - digital artifacts that matter the most in university research. More granular information is also available: the team can generate additional git tags for developments they deem relevant.

% This section does not bring much. TM
%\input{sections/build-system-short}

\subsection{Cross-linking publications, secondary data, and software archives} 
\label{ssec:cross-linking}

\begin{figure}[tb]
    \centering
    \includegraphics[width=\linewidth]{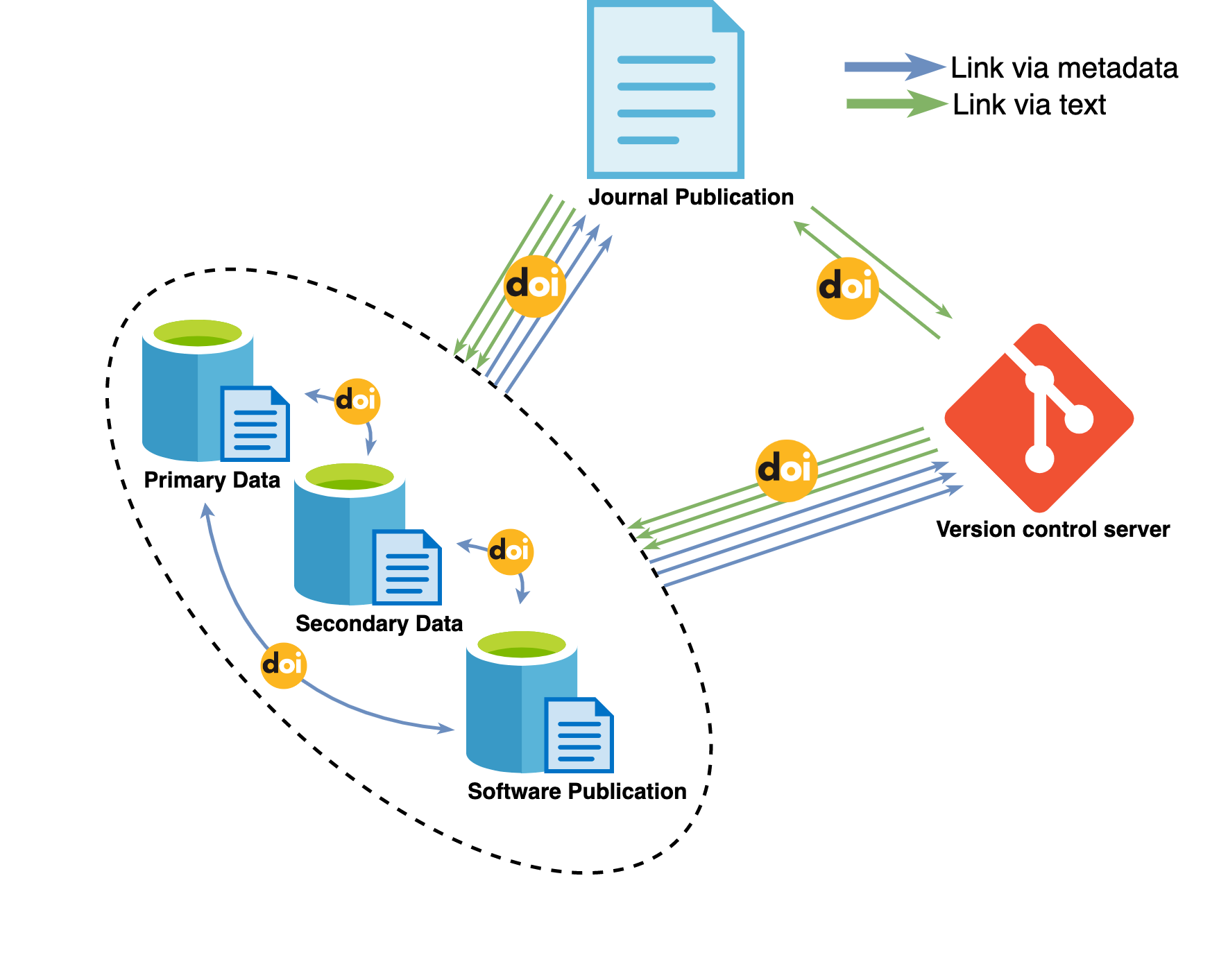}
    \caption{Cross-linked publication, data and source code using Persistent Identifiers.}
    \label{fig:cross}
\end{figure}

% \comment[id=tm]{Remove Figure 4}

In the previous section, we have outlined the process of creating PIDs for the source code, related data and reports for project milestones. We have also already introduced three links: the source code publication links to a corresponding \textit{git tag} in the code repository, the code repository states the PIDs of all the related published datasets in the \textit{README} file, and the report contains the PIDs in its text.

To enable finding all other research artifacts after finding a single artifact, we suggest cross-linking all PIDs in the metadata of all datasets (cf. \cref{fig:cross}). The metadata of a dataset can be modified without generating a new version (i.e.\ a new PID); therefore, one can cross-link them after their creation when all PIDs are available. 

Up to this point, we have not categorized research data. Here we further categorize CSE research data into configuration and input data, as well as \textit{primary} and \textit{secondary} output data. Configuration and input data are necessary for the research software to generate output. Input data should be part of the code repository and code publication unless the size of the input data prohibits this, in which case it should be archived in a data repository and fetched when needed.

Scientific reports typically discuss \textit{secondary data} in tables and diagrams, which are much smaller than \textit{primary data}. As part of the \minwf, we recommend that secondary data is published and cross-linked to facilitate direct result comparison without the error-prone, time-consuming, and often impossible re-digitalization of secondary data from publications. Scripts to produce the \textit{secondary data} from \textit{primary data} should be part of the code repository and publication, making it possible to (re-)produce the relevant tables and figures automatically from primary data. 
%This process can be very compute-intensive and time-consuming, and in full workflow we will discuss further extensions (see~\cref{TODO}).

The cross-linking of secondary data with the publication and the source code requires minimal effort compared to the overall research and publication process but delivers immediate and significant benefits. Available secondary data can be easily reused by peers for method comparisons, thereby increasing the impact of the publication. An analysis of secondary data long after the publication acceptance is possible, potentially leading to new ideas. Moreover, it becomes possible to perform regression tests that ensure new methods are at least as good as existing published methods for a specific software version. 
\subsection{FULL WORKFLOW}
\label{sec:full-workflow-short}

% This section does not bring much.
%\input{sections/issue-tracking-short}

\subsection{Test-Driven Development for Research Software}
\label{ssec:tdd}

\begin{figure*}[tb]
    \centering
    \includegraphics[width=\linewidth]{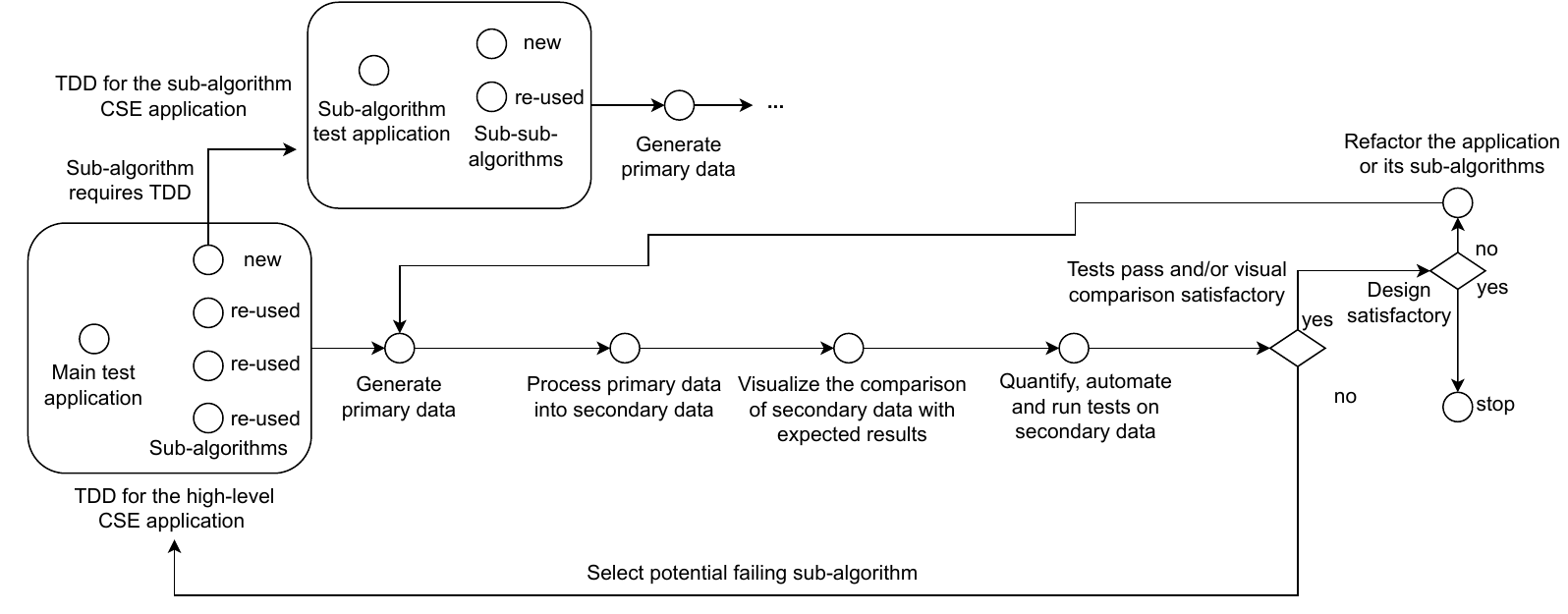}
    \caption{Test-Driven Development (TDD) applied to CSE research software in a top-down approach, starting with the top-level CSE application and recursively testing its new or re-used sub-algorithms recursively, on a per-need basis.}
    \label{fig:rsetdd}
\end{figure*}

Test-Driven Development (TDD\footnote{Kent Beck. Test-driven development: by example.
Addison-Wesley Professional, 2003}) is the practice of writing test applications before implementing the underlying algorithms. 
Test-driven development places the programmers immediately in the role of users of their own software, which, in turn, motivates a cleaner implementation\footnote{Martin, Robert C. Clean code: a handbook of agile software craftsmanship. Pearson Education, 2009.}, i.e. a cleaner Application Programming Interface (API). The TDD approach consists of three phases: red, green and refactoring. The red phase begins by implementing tests that immediately fail because of the missing or incomplete algorithmic implementation. This first step defines the  API since the tests define the function names, their return types, and arguments. The algorithm implementation follows until the tests pass, entering the green phase of TDD. In the final refactoring phase, the passing tests allow the developers to refactor the API and its underlying algorithms, using the tests to verify the correctness of the implementation. Refactoring continues until the modularity of the implementation is satisfactory.

We propose an adaptation of TDD to research software in CSE, visualized in~\cref{fig:rsetdd}. High-level CSE applications consist of complex sub-algorithms that require extensive and automatic verification and validation. Some algorithms must be programmed anew; some are re-used from legacy code. We suggest implementing the high-level CSE application first (as a high-level \textit{test}), consisting of calls to required sub-algorithms and any data processing steps. Verification and validation tests from previous publications provide a basis for comparing existing and new methods. If no previous work is available, method comparison relies on manufactured solutions (verification) and experimental data (validation). Writing CSE applications first defines the type and structure of primary data (e.g., velocity, pressure, and temperature fields) and expected macroscopic quantities (e.g. drag resistance) as shown in \cref{fig:rsetdd}. In the red phase, without available algorithmic implementation, artificial unsatisfactory results are used (hardcoded) to enable result comparison and visualization, i.e. lay out the test structure and result comparison for the initial failing tests. Similarly, failing re-used algorithms result in a failing error quantification.

With the test structure prepared, the algorithmic implementation follows, either by re-using existing algorithms or, if necessary, implementing new ones. Our version of TDD recursively branches off into sub-cycles for failing algorithms (new or re-used) until all algorithms used in the high-level CSE application work and are refactored (see~\cref{fig:rsetdd}).  

% \comment[id=dg]{Tomislav, the stuff below I don't really "understand" or at least the argument is not obvious to me...}
% \comment[id=tm]{Re-written the text ABOVE, and BELOW, let me know if it is more clear.}

Automatic testing and visualization of results in the red phase help the programmers isolate troublesome sub-algorithms more easily. Finalizing the green phase for the high-level application does not require unit testing of \emph{every} sub-algorithm in a CSE application. Covering a large-scale legacy research software that lacks unit-tests with unit-tests is impossible within a PhD research project. Our approach tests and improves upon only those new and re-used sub-algorithms (unit-tests included), that are used in scientific publications of the research group.

% \comment[id=dg]{This I also don't really get. Shouldn't each publication yield high test coverage already? Why does the overall test coverage increase?}
% \comment[id=tm]{Changed it a bit, with all the paragraphs now changed, I hope it is clear, if not, I'll do another run at it.}

As work continues on scientific publications, the test suite for the whole research software grows, with new tests resulting from each publication. Therefore, the proposed TDD approach increases overall test coverage only along the research roadmap and integrates changes using the proposed version-control branching model (cf. \nameref{ssec:vcs}). This reduces the work overhead of the researchers related to software testing.

\subsection{Test quantification and visualization}
\label{ssec:tests}

\begin{figure*}[tb]
    \centering
    \includegraphics[width=0.95\textwidth]{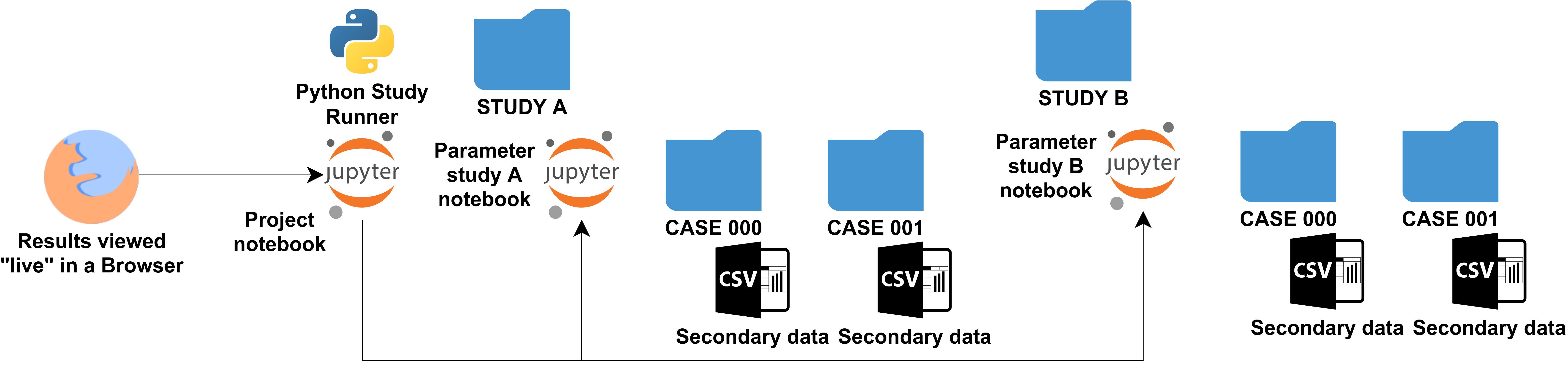}
    \caption{Different parameter studies are orchestrated by a Study Runner. All studies are implemented using Jupyter notebooks, which contain a problem description and the visualization of results. This allows for \emph{live} inspection of test result data. Structured metadata is exported into files to be available for subsequent use in publications and data cross-linking.}
    \label{fig:testvis}
\end{figure*}

Testing scientific software involves running parameter variations (studies) as illustrated in \cref{fig:testvis}.
The generally large number of test cases requires quick identification of failing tests and the reason behind the failure, as well as the ability to re-run the cases. 
Input parameters are varied in form of \emph{parameter vectors} by a \emph{study runner} on an HPC cluster.
Each parameter variation stores a map from each simulation case, identified by a case-ID local to the results folder, to the particular input parameter vector used.
Moreover, it exports structured metadata, used subsequently for cross-linking artifacts, as described in \nameref{ssec:cross-linking}.
\Cref{fig:testvis} shows a possible organization of the data in folders with a naming scheme of simple ascending integers used as locally unique IDs.

Examining test results early is important to stop simulations and save compute resources, improve job priority on the HPC cluster and speed up research. Manual inspection of tables and diagrams is a natural part of the research process and Jupyter notebooks are a solution for analyzing test results manually and in real-time. A Jupyter notebook is written for each individual study and links all the other notebooks to effectively link all tests used to verify/validate the scientific software in one place.

Each Jupyter notebook documents its parameter study, including the mathematical description of the Initial/Boundary Value Problem. 
The notebooks process secondary data generated by the simulation and store them as CSV files. 
The central project notebook can be viewed live, making it possible to check simulation results of many studies and stop them early if necessary. 
Additionally, since Jupyter notebooks are used for automatic processing of secondary data they use as input, they are always executed anew, and do not require information on the sequence of execution of their cells (provenance).
An important benefit of this visualization process is that diagrams and related secondary data processed by the notebooks can be used directly in the scientific publication.

\paragraph{Interoperable secondary data}
If peers want to reuse published secondary data, for instance,
to perform different analyses on it, they have to understand the underlying
data structure. Oftentimes, file formats that can represent nested data structures,
as \eg \textit{JSON} or \textit{HDF5}\footnote{Koranne, Sandeep, and Sandeep Koranne. "Hierarchical data format 5: HDF5." Handbook of open source tools (2011): 191-200.},
are chosen because they also provide an elegant way to store
metadata alongside the actual data. However, the structure of such files can
become arbitrarily complex, thereby making the reuse more difficult. In our workflow,
we have adopted a pragmatic solution at the cost of a small storage overhead: storing secondary data in an ASCII CSV format and saving metadata alongside the result data directly in the columns. This makes the data completely inter-operable and easy to include in data analysis. However, it does introduce repetition of metadata.
Not relying on complex file formats simplifies the processing of secondary data immensely and saves time, with a negligible storage overhead caused by (meta)data duplication.
A concrete example of storing metadata together with secondary data in tabular form is shown for hyperparameter tuning of an artificial neural network (NN) in \cref{table:metadata}. 
Fixing one metadata parameter and varying others (present in many types of parameter space explorations) makes any repetition of the fixed parameter redundant. In an attempt to save storage space, a structure of the metadata is generally defined and metadata is stored (literally or conceptually) separately from data in hierarchical data formats, or hierarchical folder structures. All those approaches significantly and unnecessarily complicate the understanding, interoperability, and analysis of secondary data. We propose to store secondary data from scientific publications in a simple tabular form. In each row of the table, a metadata set is defined by a sub-set of columns (e.g., identified by the prefix "PARAM\_" in \cref{table:metadata}). A data set in each row of the table is constructed from columns without the "PARAM\_" prefix. This way, a metadata set in a row uniquely defines a data set in the same row. Since secondary data is very small, \replaced[id=tm]{this duplication of}{duplicating metadata} information is a negligible price to pay, compared to the benefits it brings. \added[id=tm]{The straightforward CSV tabular format for secondary data and metadata significantly simplifies result comparison and data analysis}.

%Duplicating metadata\added[id=tm]{to uniquely categorize secondary data by simply repeating metadata in every row of a table that contains secondary data}\deleted[id=tm]{, which} can be ignored for secondary data that is typically small in size. \added[id=tm]{The result is a table that contains secondary data and metadata that identifies it uniquelly, regardless of the parameter exploration approach used to generate it. Each row in the table contains two vectors: a metadata vector identified by columns with the prefix "_PARAM", and a data vector.}

\begin{table*}[!htb]
    \scriptsize
    \centering
    \begin{tabular}{llrrrrr}
\toprule
{} PARAM\_HIDDEN\_LAYERS &  PARAM\_OPTIMIZER\_STEP &  PARAM\_MAX\_ITERATIONS &   EPOCH & TRAINING\_MSE \\
\midrule
10,10,10,10 &          0.0001 &            3000 &      1 &        1.091560 \\
10,10,10,10 &          0.0001 &            3000  &      2 &        1.082970 \\
10,10,10,10 &          0.0001 &            3000  &      3 &        1.077200 \\
10,10,10,10 &          0.0001 &            3000 &      4 &        1.072650 \\
...         &          ...    &            ...  &   ...    &    ... &        ... \\
10,10,10,10 &           0.001 &            3000  &      1 &        0.992354 \\
10,10,10,10 &           0.001 &            3000 &      2 &        0.991959 \\
10,10,10,10 &           0.001 &            3000 &      3 &        0.995102 \\
10,10,10,10 &           0.001 &            3000 &      4 &        0.996143 \\
...         &          ...    &            ...  &   ...    &    ... &        ... \\
\bottomrule
\end{tabular}

    \medskip
    \caption{Storing metadata together with secondary data in a tabular form repeats metadata (columns with 'PARAM\_' prefix); however, it makes further data analysis of secondary data straightforward, and this CSV format completely interoperable.}
    \label{table:metadata}
\end{table*}

\subsection{Containerization}
\label{ssec:containerization}

One major obstacle to the reproducibility and reusability of research software lies in its dependencies, that is, other software needed for the research software to run. In order for peers to reuse the software of others, all its dependencies must be publicly accessible. Dependency installation is generally tricky and time-consuming. It requires an up-to-date and accurate documentation about the dependencies, i.e. their exact versions and compilation options.

A widely used approach for encapsulating software dependencies is containerization, and Docker\footnote{\url{https://docs.docker.com/get-started/overview}} and Singularity are two widespread containerization software solutions. Containerization software encapsulates software dependencies in so-called images, which can be used to spin up containers to run the research software in. Therefore, containerization software replaces all other dependencies, making it significantly more straightforward to reuse the software, explore and modify it and reproduce results. 

So-called \emph{recipe} files determine the ingredients (dependencies and their installation instructions) in a container image, starting from a \textit{base image} that typically represents the layer of the operating system. Besides being machine-readable and executable, a containerization \emph{recipe} is also a valuable source of documentation on how to recover the suitable environment of the research software.

Large input data should not be included in the image; instead, large data should be published separately and made uniquely identifiable and accessible (cf. 
 \nameref{ssec:cross-linking}). This approach keeps the container size as small as possible. For instance, for users that want to use the software on their own data set. Large input data can be downloaded from the data repository to reproduce results if needed.

To make \emph{recipe} files as \emph{FAIR} as possible, we suggest retrieving all data required to build the image from persistent online resources. For example, any \textit{copy} statement in the build recipe that copies files from a specific computer we use to build the image (i.e. the host machine) into the image, uses paths and data only available on that host machine. Moreover, these statements do not carry any notion of versioning, and copying the wrong version of the files can cause various errors during the build process. Note that git repositories are not persistent: they do not guarantee long-term accessibility or permanence. Therefore, if available, one can obtain research software in the \emph{recipe} from \href{https://www.softwareheritage.org/}{softwareheritage.org/}, which guarantees availability over extended time frames.

Usually, one can define a set of commands to be executed when an image is used to spin up a container. This can be used to automatically invoke the commands required to (re-)produce the results discussed in a journal publication. However, running the software on a different machine inside a container is likely to produce slightly different results than the published ones: bit-wise reproducibility is generally unattainable. Therefore, successful reproduction of results should include tolerable deviations that can be checked using tools like fieldcompare~\citep{Glaeser2023}.

Finally, note that additional effort and modifications are required to ensure the images (containers) reproduce CSE results on different high-performance-computing (HPC) systems with similar computational efficiency. The reason behind this is the necessity to use HPC frameworks for parallel communication that were specifically tuned to an HPC system. Our workflow does not utilise containerization for \emph{automatic} and computationally efficient reproducibility on HPC systems.

\subsection{Continuous integration of research software}
\label{ssec:continuous-integration}

\begin{figure*}[!ht]
    \centering
    \includegraphics[scale=0.5]{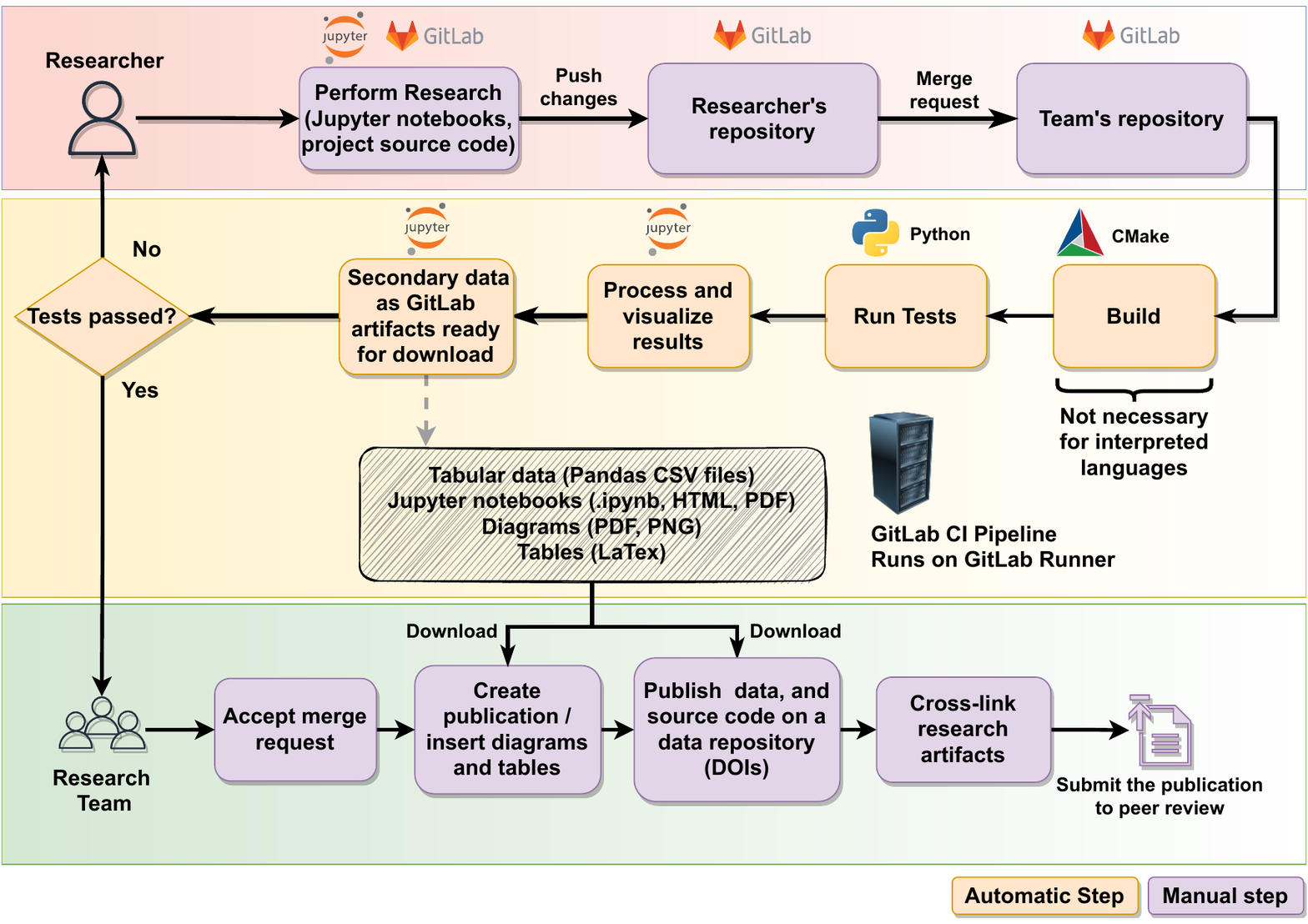}
    \caption{Continuous integration workflow: the merge request triggers the CI pipeline to build the software, run all tests and visualize the results. If tests pass, the merge request is accepted and the changes are added to the main repository. If the tests fail, the researcher can inspect the published visualization of results.}
    \label{fig:ci_cd}
\end{figure*}

Continuous Integration (CI)\footnote{Martin Fowler and Matthew Foemmel. Continuous
integration, 2006.} is the practice of frequent integration of changes into a single version of the software. To understand CI, it is beneficial to briefly revisit the related branching model from the section \nameref{ssec:vcs}. A new research idea is developed using a so-called feature branch, while the current stable version of the software is maintained on the main branch. In traditional software engineering, CI recommends frequent, daily integration from features onto the main branch, but for scientific software this is not always possible as the features are driven by scientific research that requires much more time for incremental improvements. 
%To \replaced[id=tm]{prevent}{anticipate} diverging versions with \replaced[id=tm]{common improvements}{individual feature branches}, integration of improved functionality shared across different research ideas should be done as \replaced[id=tm]{soon as possible (e.g. as soon as the functionality is in the green TDD phase \cref{ssec:tdd})}{frequently as possible, e.g., weekly.} \comment[id=dg]{I think in the following sentence something is wrong}\replaced[id=tm]{This ensures that improvements shared across different research ideas are integrated and do not cause complicated conflicts, or duplicated functionality.}{This ensures that improvements to common, shared components are integrated into the individual feature branches, and that developments achieved in a successful project are integrated by everyone in the research group.}
% \comment[id=tm]{See commented out text. If we need to say something about the branching model and integration frequency, we should do this in the branching-model section, right?}
% \comment[id=dg]{@tomislav, from my side this is fine. In the section on VC it is mentioned that after the team has checked everything (e.g. upon submission), the branch can be merged although the review is still ongoing - in order to have the improvements in main. From my side this is enough...}

\Cref{fig:ci_cd} contains a schematic depiction of the CI workflow for research software.
Feature development happens in the user's repository. The user opens a merge request when the feature is implemented together with its tests, triggering the CI pipeline. The build system, CMake in this example, checks the software's compilation on different platforms. If the build tests pass, the error-quantification tests are executed on the GitLab runner, including unit tests and small-scale Verification\&Validation (V\&V) parameter studies. These tests also populate the folder structure with secondary data as depicted in~\cref{fig:testvis}.

The automatic execution of tests as part of the CI pipeline is an important aspect of software quality assurance, and simplifies the integration of new versions.
A merge request should only be accepted if all tests have passed.
This ensures that the integration of all changes, i.e., feature branch and master branch, do not lead to errors.

Before the quantification of tests causes tests to either fail or pass, the results are processed to show \emph{why} the tests failed / succeeded.
Generally, the simple fail/pass output of the test environment is not sufficient to diagnose test failure.
Especially in the CSE context, tests are complex and require documentation, and test results require visualization in the form of diagrams or $2$D/$3$D visualization to determine the cause of an error.
For this purpose, the proposed CI pipeline from \cref{fig:ci_cd} contains jobs that export Jupyter notebooks that document the tests and visualize the results as HTML. The final result is a complete documentation and visualization of CSE tests and their results, uniquely identified with a commit.

The test suite in this workflow includes tests generated by CSE-oriented Test Driven Development, ranging from unit-tests to verification and validation tests. Validation and verification tests in the CI pipeline are exactly the same as large-scale V\&V tests used in the publications; however, they are scaled down using smaller data (e.g. resolution of a PDE discretization, a smaller neural network) so that they can be executed on many-core workstations, instead of large HPC systems. This way, the CI documents the computational workflows necessary to reproduce large-scale scientific results. An actual reproduction can be done by simply scaling up the data (resolution) of the V\&V tests taken from the CI pipeline, and run them manually on an HPC cluster. We do not utilise CI for reproducing large-scale validation tests on purpose, because this requires IT infrastructure and teams responsible for installing and maintaining a secure CI service on a University HPC cluster, which is not available to all research groups. Furthermore, adding large-scale reproduction tests to the CI complicates it significantly in the long-term as the project grows. Large-scale tests require time and resources, naturally leading to selecting a sub-set of large-scale test to run within the CI pipeline. Large-scale tests are prone to failure and may require restarts, further complicating CI implementation on HPC systems. Adding large-scale tests to CI is challenging in international collaborations, where adaptations must be made to containers for efficient execution on different HPC systems. There are questions on security, that arise when using CI on HPC clusters. Finally, modern many-core workstations are affordable, and powerful enough for running V\&V tests whose data resolution ensures sufficient accuracy for identifying computational errors. We therefore propose using CI for small to mid-scale V\&V tests, documenting the software dependencies, while reproducing large-scale results on HPC systems outside of CI.

The test pipeline in \cref{fig:ci_cd} visually shows the relationships between target quantities and parameters, allowing developers to quickly identify failures. Regression tests are emphasized, in which newly produced results are compared with reference solutions to detect any changes in the software's behavior \citep{Glaeser2023}.

Provided all tests pass, the acceptance of the MR in \cref{fig:ci_cd} is decided by the responsible team member, who can also provide feedback on code quality.
A web interface to git, such as GitLab, simplifies the code-review process within the research group.

\subsection{Cross-linking with software images and primary data}

The \minwf{} cross-linking links source code, publication, and secondary data. For the full workflow, we recommend to extend it with primary data and containers, enabling readers to gain further insights and making results easier to reproduce across platforms. Additional steps are covered in \cref{fig:workflow-overview}.

The primary data is archived differently due to its size, with each parameter study stored separately for easier access and to save time and network bandwidth. The directory structure is similar to the one for secondary data, but each study is stored in a separate archive. Subsets of the results, such as "STUDY A" or "STUDY B," can be accessed with their Jupyter notebooks, HTML exports, and parameter study information. Generated datasets for subsets of primary data are uploaded to a data repository and linked using PIDs as described in \minwf{}.

Containers (cf. section \nameref{ssec:containerization}) should be archived alongside the source code publication. This makes it possible for other researchers to use and explore the software inside the container without going through the installation process. In our workflow, we use Singularity containers supported by our HPC center, which are single files that can be easily uploaded to a data repository. But other container systems are also possible, for instance, Docker allows users to export images to compressed archives from which the images can be reloaded at a later time.

It is important to publish the recipe for building the image along with the actual image. For instance, images built on and for a specific architecture may not work on another architecture. With the availability of the recipe, users can build the container natively on their system. Finally, if that also ceases to work over time, the recipe documents the environment of the researcher who produced the results, which can help users to reinstantiate a suitable environment in the future. To ensure consistency over time, the recipe should use a version-pinned base image and install pinned versions of dependencies.

\section{Minimal working example}
\label{sec:minimal-example}

The Minimal Working Example\footnote{\url{https://tudatalib.ulb.tu-darmstadt.de/handle/tudatalib/3522}} (MWE) serves as a basis for the minimal version of the proposed workflow, containing a version-controlled open-source C++ application generating a small data set, visualized in a Jupyter notebook, and referenced in a scientific report. The MWE software is version-controlled and built using the CMake build system. The example data sets, the repository snapshot, the active repository, and the minimal report are all cross-linked.

% TMI
%The MWE research software is a simple C++ application that evaluates numerical derivatives of a polynomial using finite differences in a 1D interval. The finite differences are compared against exact values for a prescribed polynomial. 

The MWE is archived at the \href{https://tudatalib.ulb.tu-darmstadt.de/handle/tudatalib/3522}{\textcolor{blue}{TUdatalib data repository}}. The research data and the snapshot of the research code  are archived, and cited in the example Research Report, along with the URL of the source code repository. Once the research report is uploaded to the repository, the \emph{metadata} of the other data items are updated by adding the 'dc.relation.isreferencedby' elements to their metadata together with the report's DOI, thereby cross-linking the report with the data items. This way, the \href{https://tudatalib.ulb.tu-darmstadt.de/handle/tudatalib/3530?show=full}{\textcolor{blue}{full metadata record}} of a data item on the repository stores the cross-linking information, that can be updated as data-items evolve. For example, as the peer-review process progresses or new milestones are reached, new sub-versions of data items obtain new DOIs that denote sub-versions, e.g. \url{https://doi.org/10.48328/tudatalib-921.2} for a new version of the Research Report. This makes it possible to continue updating the Report, Data, and Code Snapshot with the feedback from the peer-review process or during the development of new and improved methods.

A minimal CI pipeline is configured for the GitLab MWE source code repository, demonstrating the use of Jupyter notebooks for data processing and visualization. The research data linked with the report is the artifact generated by the CI pipeline. For realistic CSE tests, Jupyter notebooks contain detailed information about the test setup: geometry of the problem, initial and boundary conditions, model parameters.

Case studies with real-world research projects are described in \citep{Maric2022rse}. 

\section{Conclusions}
\label{sec:conclusions}

We propose a Research Software Engineering workflow for Computational Science and Engineering, that focuses on improving research software in the context of scientific publications, the most valued research artifacts in a university setting. This pragmatic approach introduces a minimal work overhead and is suitable for university research groups without resources dedicated to software engineering. We propose a branching model for version control tailored to the peer-review process and an adaptation of Test-Driven Development to research software using Validation and Verification tests from scientific publications. We recommend to use a simple, open and inter-operable format for secondary data (diagrams and tables) and their metadata, continuous integration for automating and documenting scientific workflows with unit-tests and low-resolution Validation\&Verification tests, and cross-linking all research artifacts using PIDs. This increases the overall quality of the scientific output, and simplifies the combination and re-use of research ideas embodied in research software.

\section{ACKNOWLEDGMENTS}

This work was funded by the Deutsche Forschungsgemeinschaft (DFG, German Research Foundation) -- Project-ID 265191195 -- SFB 1194, and by the Hessian LOEWE initiative within the Software-Factory 4.0 project.
The authors would also like to thank the Federal Government and the Heads of Government of the Länder, as well as the Joint Science
Conference (GWK), for their funding and support within the framework of the NFDI4Ing consortium. Funded by the German Research
Foundation (DFG) - project number 442146713.
Benchmarks were conducted on the Lichtenberg cluster at TU Darmstadt.

The authors are grateful to Moritz Schwarzmeier for his work on the NFDI4Ing Knowledgebase, and to Moritz Schwarzmeier and Tobias Tolle for providing valuable constructive feedback on the RSE workflow and this manuscript.

\def\refname{REFERENCES}

\bibliographystyle{abbrvnat} 
\bibliography{literature.bib}

\begin{thebibliography}{12}
\providecommand{\natexlab}[1]{#1}
\providecommand{\url}[1]{\texttt{#1}}
\expandafter\ifx\csname urlstyle\endcsname\relax
  \providecommand{\doi}[1]{doi: #1}\else
  \providecommand{\doi}{doi: \begingroup \urlstyle{rm}\Url}\fi

\bibitem[FAI(2022)]{FAIR}
{FAIR principles}.
\newblock \url{https://www.go-fair.org/fair-principles/}, 2022.
\newblock Accessed: 2022-08-03.

\bibitem[Anzt et~al.(2019)Anzt, Chen, Cojean, Dongarra, Flegar, Nayak,
  Quintana-Ortí, Tsai, and Wang]{anzt_towards_2019}
H.~Anzt, Y.-C. Chen, T.~Cojean, J.~Dongarra, G.~Flegar, P.~Nayak, E.~S.
  Quintana-Ortí, Y.~M. Tsai, and W.~Wang.
\newblock Towards {Continuous} {Benchmarking}: {An} {Automated} {Performance}
  {Evaluation} {Framework} for {High} {Performance} {Software}.
\newblock In \emph{Proceedings of the {Platform} for {Advanced} {Scientific}
  {Computing} {Conference}}, {PASC} '19, pages 1--11, New York, NY, USA, June
  2019. Association for Computing Machinery.
\newblock ISBN 978-1-4503-6770-7.
\newblock \doi{10.1145/3324989.3325719}.
\newblock URL \url{https://doi.org/10.1145/3324989.3325719}.

\bibitem[Fehr et~al.(2016)Fehr, Heiland, Himpe, and Saak]{Fehr2016}
J.~Fehr, J.~Heiland, C.~Himpe, and J.~Saak.
\newblock Best practices for replicability, reproducibility and reusability of
  computer-based experiments exemplified by model reduction software.
\newblock \emph{AIMS Mathematics}, 1\penalty0 (3):\penalty0 261--281, 2016.
\newblock ISSN 24736988.
\newblock \doi{10.3934/Math.2016.3.261}.
\newblock arXiv: 1607.01191.

\bibitem[Gläser et~al.(2023)Gläser, Koch, Peters, Marcus, and
  Flemisch]{Glaeser2023}
D.~Gläser, T.~Koch, S.~Peters, S.~Marcus, and B.~Flemisch.
\newblock fieldcompare: A python package for regression testing simulation
  results.
\newblock \emph{Journal of Open Source Software}, 8\penalty0 (81):\penalty0
  4905, 2023.
\newblock \doi{10.21105/joss.04905}.
\newblock URL \url{https://doi.org/10.21105/joss.04905}.

\bibitem[Heroux et~al.(2020)Heroux, Gonsiorowski, Gupta, Milewicz, Moulton,
  Watson, Willenbring, Zamora, and Raybourn]{2020:heroux:psip}
M.~A. Heroux, E.~Gonsiorowski, R.~Gupta, R.~Milewicz, J.~D. Moulton, G.~R.
  Watson, J.~Willenbring, R.~J. Zamora, and E.~M. Raybourn.
\newblock Lightweight software process improvement using productivity and
  sustainability improvement planning (psip).
\newblock In G.~Juckeland and S.~Chandrasekaran, editors, \emph{Tools and
  Techniques for High Performance Computing}, pages 98--110, Cham, 2020.
  Springer International Publishing.
\newblock ISBN 978-3-030-44728-1.

\bibitem[Jiménez et~al.(2017)Jiménez, Kuzak, Alhamdoosh, Barker, Batut, Borg,
  Capella-Gutierrez, Chue~Hong, Cook, Corpas, Flannery, Garcia, Gelpí,
  Gladman, Goble, González~Ferreiro, Gonzalez-Beltran, Griffin, Grüning,
  Hagberg, Holub, Hooft, Ison, Katz, Leskošek, López~Gómez, Oliveira,
  Mellor, Mosbergen, Mulder, Perez-Riverol, Pergl, Pichler, Pope, Sanz,
  Schneider, Stodden, Suchecki, Svobodová~Vařeková, Talvik, Todorov,
  Treloar, Tyagi, van Gompel, Vaughan, Via, Wang, Watson-Haigh, and
  Crouch]{Jimenez2017}
R.~C. Jiménez, M.~Kuzak, M.~Alhamdoosh, M.~Barker, B.~Batut, M.~Borg,
  S.~Capella-Gutierrez, N.~Chue~Hong, M.~Cook, M.~Corpas, M.~Flannery,
  L.~Garcia, J.~L. Gelpí, S.~Gladman, C.~Goble, M.~González~Ferreiro,
  A.~Gonzalez-Beltran, P.~C. Griffin, B.~Grüning, J.~Hagberg, P.~Holub,
  R.~Hooft, J.~Ison, D.~S. Katz, B.~Leskošek, F.~López~Gómez, L.~J.
  Oliveira, D.~Mellor, R.~Mosbergen, N.~Mulder, Y.~Perez-Riverol, R.~Pergl,
  H.~Pichler, B.~Pope, F.~Sanz, M.~V. Schneider, V.~Stodden, R.~Suchecki,
  R.~Svobodová~Vařeková, H.~A. Talvik, I.~Todorov, A.~Treloar, S.~Tyagi,
  M.~van Gompel, D.~Vaughan, A.~Via, X.~Wang, N.~S. Watson-Haigh, and
  S.~Crouch.
\newblock Four simple recommendations to encourage best practices in research
  software.
\newblock \emph{F1000Research}, 6:\penalty0 1--15, 2017.
\newblock ISSN 1759796X.
\newblock \doi{10.12688/f1000research.11407.1}.

\bibitem[Maric et~al.(2022)Maric, Gl{\"a}ser, Lehr, Papagiannidis, Lambie,
  Bischof, and Bothe]{Maric2022rse}
T.~Maric, D.~Gl{\"a}ser, J.-P. Lehr, I.~Papagiannidis, B.~Lambie, C.~Bischof,
  and D.~Bothe.
\newblock A research software engineering workflow for computational science
  and engineering.
\newblock \emph{arXiv preprint arXiv:2208.07460}, 2022.

\bibitem[Riesch et~al.(2020)Riesch, Nguyen, and Jirauschek]{riesch_bertha_2020}
M.~Riesch, T.~D. Nguyen, and C.~Jirauschek.
\newblock bertha: {Project} skeleton for scientific software.
\newblock \emph{PLOS ONE}, 15\penalty0 (3):\penalty0 e0230557, Mar. 2020.
\newblock ISSN 1932-6203.
\newblock \doi{10.1371/journal.pone.0230557}.
\newblock URL
  \url{https://journals.plos.org/plosone/article?id=10.1371/journal.pone.0230557}.
\newblock Publisher: Public Library of Science.

\bibitem[Sampedro et~al.(2018)Sampedro, Holt, and
  Hauser]{sampedro_continuous_2018}
Z.~Sampedro, A.~Holt, and T.~Hauser.
\newblock Continuous {Integration} and {Delivery} for {HPC}: {Using}
  {Singularity} and {Jenkins}.
\newblock In \emph{Proceedings of the {Practice} and {Experience} on {Advanced}
  {Research} {Computing}}, {PEARC} '18, pages 1--6, New York, NY, USA, July
  2018. Association for Computing Machinery.
\newblock ISBN 978-1-4503-6446-1.
\newblock \doi{10.1145/3219104.3219147}.
\newblock URL \url{https://doi.org/10.1145/3219104.3219147}.

\bibitem[Stanisic et~al.(2015)Stanisic, Legrand, and
  Danjean]{stanisic_effective_2015}
L.~Stanisic, A.~Legrand, and V.~Danjean.
\newblock An {Effective} {Git} {And} {Org}-{Mode} {Based} {Workflow} {For}
  {Reproducible} {Research}.
\newblock \emph{ACM SIGOPS Operating Systems Review}, 49\penalty0 (1):\penalty0
  61--70, Jan. 2015.
\newblock ISSN 0163-5980.
\newblock \doi{10.1145/2723872.2723881}.
\newblock URL \url{https://doi.org/10.1145/2723872.2723881}.

\bibitem[Wilson et~al.(2014)Wilson, Aruliah, Brown, Chue~Hong, Davis, Guy,
  Haddock, Huff, Mitchell, Plumbley, Waugh, White, and Wilson]{Wilson2014}
G.~Wilson, D.~A. Aruliah, C.~T. Brown, N.~P. Chue~Hong, M.~Davis, R.~T. Guy,
  S.~H. Haddock, K.~D. Huff, I.~M. Mitchell, M.~D. Plumbley, B.~Waugh, E.~P.
  White, and P.~Wilson.
\newblock Best {Practices} for {Scientific} {Computing}.
\newblock \emph{PLoS Biology}, 12\penalty0 (1), 2014.
\newblock ISSN 15457885.
\newblock \doi{10.1371/journal.pbio.1001745}.
\newblock arXiv: 1210.0530.

\bibitem[Wilson et~al.(2017)Wilson, Bryan, Cranston, Kitzes, Nederbragt, and
  Teal]{wilson_good_2017}
G.~Wilson, J.~Bryan, K.~Cranston, J.~Kitzes, L.~Nederbragt, and T.~K. Teal.
\newblock Good enough practices in scientific computing.
\newblock \emph{PLOS Computational Biology}, 13\penalty0 (6):\penalty0
  e1005510, June 2017.
\newblock ISSN 1553-7358.
\newblock \doi{10.1371/journal.pcbi.1005510}.
\newblock URL
  \url{https://journals.plos.org/ploscompbiol/article?id=10.1371/journal.pcbi.1005510}.
\newblock Publisher: Public Library of Science.

\end{thebibliography}

\end{document}